\documentclass[final,5p,times,twocolumn]{elsarticle}

\journal{Surface Science}

\usepackage{amssymb}
\usepackage{amsmath}
\usepackage{graphicx}
\usepackage{booktabs}
\usepackage{xcolor}
\usepackage{xr}
\usepackage{hyperref}
\usepackage{multirow}
\usepackage{array}
\usepackage{float}

\begin{document}


\begin{frontmatter}

\title{Improved screening of doped ZrO$_2$ catalysts for CO$_2$ hydrogenation from graph-based machine learning}

\author[aff1]{Karsten Walz Jensen}
\author[aff1]{Mie Andersen\corref{cor1}}
\ead{mie@phys.au.dk}
\author[aff1]{Raffaele Cheula\corref{cor1}}
\ead{raffaele.cheula@phys.au.dk}

\cortext[cor1]{Corresponding authors}

\affiliation[aff1]{
    organization={Center for Interstellar Catalysis, Department of Physics and Astronomy},
    addressline={Aarhus University},
    city={Aarhus C},
    postcode={8000},
    country={Denmark}
}

\begin{abstract}
Dopants can tune the activity of zirconia-based CO$_2$ hydrogenation 
catalysts through changes in adsorbate and transition-state (TS) energetics. Computing 
these quantities with density functional theory (DFT) becomes expensive 
when several dopants, reaction intermediates, and TSs must be screened.
Here, we further develop the graph-based machine learning method Wasserstein Weisfeiler–Lehman Gaussian Process Regression (WWL-GPR): we introduce additional electronic-structure and local-environment features relevant for oxide surfaces and use the resulting model to predict adsorption and TS energies on 
ZrO$_2$(101) surfaces doped with Al$^{3+}$, Cd$^{2+}$, Ga$^{3+}$, In$^{3+}$, and Zn$^{2+}$.
Adsorbate-surface structures are represented as attributed 
graphs that combine bonds connectivity and 
features. We benchmark WWL-GPR against 
thermodynamic scaling relations (TSR) for reaction intermediates and 
Br\o{}nsted--Evans--Polanyi relations (BEP) for TSs, 
and then propagate the predicted energies to microkinetic 
CH$_3$OH-yield estimates. 
WWL-GPR consistently improves the prediction of reaction intermediate energetics 
relative to linear scaling approaches and provides TS estimates 
that are competitive with BEP-based models. These improvements become especially 
relevant when the predicted energetics are propagated through the reaction network.
Microkinetic calculations show that WWL-GPR-derived energies give 
CH$_3$OH yields closer to DFT-derived yields than the 
linear-relation workflow. Degree-of-rate-control analysis links the 
activity errors to dopant-dependent rate-controlling steps, 
including HCOO* formation, H$_2$COOH* dissociation, 
and H$_3$CO* formation. These results support WWL-GPR as a 
data-efficient surrogate for connecting atomistic energetics 
with catalyst-screening observables on doped oxide surfaces.
\end{abstract}




\end{frontmatter}

\section{Introduction}
\label{introduction}

Reducing anthropogenic CO$_2$ emissions while preserving 
access to carbon-based chemicals is a central challenge for a sustainable 
chemical industry. One possible route is to capture CO$_2$ from 
concentrated streams or directly from air and convert it into C1 products. 
Methanol is an attractive target because it can be used as a fuel additive, 
a hydrogen carrier, and a platform molecule for larger chemicals and olefins 
\cite{goeppert2014recycling}. Such carbon recycling is beneficial when the 
required hydrogen and process energy are supplied from low-carbon or 
renewable sources.

Industrial methanol synthesis conventionally uses \(\mathrm{Cu/ZnO/Al_2O_3}\) 
catalysts and syngas feeds containing CO, CO$_2$, and H$_2$. 
The balance between CO$_2$ hydrogenation, CO hydrogenation, and the 
reverse-water-gas-shift (RWGS) reaction controls the product distribution 
\cite{goeppert2014recycling,kattel2017active}. The three coupled reactions are
\begin{align}
\mathrm{CO_2 + 3H_2} &\rightleftharpoons \mathrm{CH_3OH + H_2O},
& \Delta H_R^{0} &= -50~\mathrm{kJ~mol^{-1}},
\label{eq:co2-hydrogenation}\\
\mathrm{CO_2 + H_2} &\rightleftharpoons \mathrm{CO + H_2O},
& \Delta H_R^{0} &= +41~\mathrm{kJ~mol^{-1}},
\label{eq:rwgs}\\
\mathrm{CO + 2H_2} &\rightleftharpoons \mathrm{CH_3OH},
& \Delta H_R^{0} &= -91~\mathrm{kJ~mol^{-1}}.
\label{eq:co-hydrogenation}
\end{align}
Equations~\eqref{eq:co2-hydrogenation} and \eqref{eq:co-hydrogenation} are exothermic, 
so higher temperature thermodynamically disfavors methanol formation. Higher temperature 
also favors the endothermic RWGS reaction in Equation~\eqref{eq:rwgs}, increasing CO and 
H$_2$O formation. Water can also compete for surface sites and 
lower reaction rates, especially on oxide catalysts 
\cite{goeppert2014recycling,cheula2024dopants}. These constraints make low-temperature 
operation and high selectivity important catalyst-design targets for methanol synthesis.

ZrO$_2$-based catalysts are promising in this context because zirconia
surfaces can promote selective CH$_3$OH-production reaction paths 
through oxygen vacancies and dopant-modified local coordination environments
\cite{cheula2024dopants,martin2016indium,wang2017znozro2,frei2019atomic,han2023cugazrox}.
Experimental studies have reported high methanol selectivity for 
doped ZrO$_2$ catalysts, with Cd$^{2+}$-, Ga$^{3+}$-, In$^{3+}$-, and Zn$^{2+}$-doped 
zirconia showing high activity, whereas other dopants, such as Al$^{3+}$, 
are substantially less active \cite{cheula2024dopants,wang2017znozro2,han2023cugazrox}. 
Cheula et al. rationalized these trends through a DFT-based microkinetic study 
of metal-doped ZrO$_2$(101), showing that catalytic activity is 
controlled by the ability of the doped surface to dissociate H$_2$ 
and to stabilize the oxygen vacancy required for methanol formation 
\cite{cheula2024dopants}. This strong sensitivity to local dopant chemistry 
is useful for catalyst optimization, but it also makes systematic screening 
challenging.

Microkinetic models are commonly used to estimate whether a candidate
surface is catalytically viable. Such models require adsorption energies
and activation barriers for the elementary steps in the reaction
network. Density functional theory (DFT) provides these quantities with
atomistic resolution, but doped-oxide screening studies rapidly become
computationally demanding when multiple dopant species and positions,
oxygen-vacancy configurations, reaction intermediates, and TSs
must be considered.
Linear scaling relations can reduce this cost:
thermodynamic scaling relations (TSR) estimate reaction-intermediate energies
from correlations within related adsorbates, while
Br\o{}nsted--Evans--Polanyi (BEP) relations connect reaction energies and
activation barriers
\cite{abildpedersen2007scaling,norskov2009towards,medford2015sabatier,evans1935applications,studt2014discovery}.
These relations are useful, but they compress local dopant effects,
adsorbate--surface coordination, and electronic-structure changes into
a small number of descriptors. As a result, linear models can sometimes
oversimplify the structure--energy relationships that determine catalytic
activity.

Machine-learning models offer a more flexible route, provided that the
representation keeps the relevant surface chemistry. Graph-based models
are attractive because adsorbate-surface systems are naturally described
by atoms and chemical bonds
\cite{butler2018machine,deringer2019machine,bartok2013representing,bartok2017unifies,xie2018crystal}.
For small and medium-sized data sets, graph kernels combined with Gaussian
process regression (GPR) are useful because they can learn from structured
inputs while also providing predictive uncertainty
\cite{rasmussen2006gpml,shervashidze2011weisfeiler,togninalli2019wasserstein}. 
For instance, the Wasserstein Weisfeiler-Lehman (WWL) graph kernel compares distributions of
continuous node embeddings using an optimal-transport distance, allowing
graphs with similar connectivity but different local chemistry to be
distinguished
\cite{shervashidze2011weisfeiler,togninalli2019wasserstein}.

Here, we build on a GPR model based on the WWL kernel that was recently developed to predict adsorption energies of complex reaction intermediates on metal alloy surfaces \cite{xu2022binding} and subsequently extended to predict also TS energies for surface reactions \cite{cheula2025}.
We further develop the WWL-GPR framework for oxide catalysis by introducing additional features that capture the local structural and electronic environment, including Bader charges, projected density-of-states features and SOAP descriptors.
The resulting model is evaluated for the prediction of reaction-intermediate and TS energetics for methanol synthesis on doped ZrO$_2$(101), considering 105 reaction intermediates and 55 TS structures across surfaces doped with Al$^{3+}$, Cd$^{2+}$, Ga$^{3+}$, In$^{3+}$, and Zn$^{2+}$.
Benchmarking against TSR and BEP relations for adsorbate and TS energy prediction is carried out using both stratified cross-validation and dopant-held-out validation, thereby assessing interpolation and transferability to unseen dopants.
Finally, we propagate the predicted energetics through a microkinetic model and show that WWL-GPR-derived energetics gives CH$_3$OH activity trends that more closely reproduce DFT-based results, while also better preserving dopant-dependent mechanistic aspects such as dominant reaction pathways and rate-controlling steps.

\section{Methods}
\label{methods}

\subsection{Doped ZrO$_2$ DFT data set}
\label{doped-zro2-data-set}

The data set utilized in this work is based on the methanol-synthesis reaction mechanism on doped
ZrO$_2$(101) surfaces reported by Cheula et al.
\cite{cheula2024dopants}. Five dopants were considered:
Al$^{3+}$, Cd$^{2+}$, Ga$^{3+}$, In$^{3+}$, and Zn$^{2+}$. 
For each doped ZrO$_2$(101) surface, the reaction mechanism includes 21 reaction
intermediates and 16 elementary steps, which describe the competing 
pathways of H$_2$COO* dissociation, H$_2$COOH* dissociation, and CO formation 
through COOH* dissociation
\cite{cheula2024dopants,martin2016indium,wang2017znozro2,frei2019atomic,studt2014discovery,graciani2014highly}.
The reaction intermediates and elementary steps are summarized in 
Table~S1 and Table~S2, respectively.
The resulting data set contains 105 reaction-intermediate structures and 55 
TS structures (activated elementary steps).

The target quantity used throughout this work is the formation energy,
$E_{\mathrm{form}}$, referenced to the corresponding clean doped
ZrO$_2$(101) surface and to gas-phase CO$_2$, H$_2$, and H$_2$O. For a
given adsorbed intermediate or TS, it is defined as
\begin{equation}
E_{\mathrm{form}} = E_{\mathrm{tot}} - E_{\mathrm{surf}} - \sum_i \nu_i E_i^{\mathrm{gas}},
\label{eq:eform}
\end{equation}
where $E_{\mathrm{tot}}$ is the DFT energy of the adsorbate--surface or
TS--surface structure, $E_{\mathrm{surf}}$ is the DFT energy of the corresponding
clean doped surface, and $E_i^{\mathrm{gas}}$ is the DFT energy of the gas-phase
reference species $i \in \{\mathrm{CO_2},\mathrm{H_2},\mathrm{H_2O}\}$.
The stoichiometric coefficients $\nu_i$ reproduce the elemental 
composition of the corresponding adsorbate or TS structure.

These DFT-derived formation energies are used as reference labels for
training and evaluating the WWL-GPR model and the linear-relation
baselines.

\subsection{Graph representation and WWL-GPR model}
\label{graph-representation-and-wwl-gpr-model}

Each atomistic structure describing an adsorbate or a TS
was encoded as an attributed graph and used as input to the WWL-GPR model to 
predict the formation energy.
In this representation, atoms are mapped to nodes
and bonds to edges. Edges were derived from the adjacency matrix describing 
the chemical bonds of the atomistic structure. The graph
construction starts from the isolated gas-phase adsorbate and the clean
doped ZrO$_2$(101) surface, for which the relevant descriptors
are computed separately. When the adsorbate is placed on the surface, the
adsorbed graph is obtained by combining the clean-surface graph with the
additional adsorbate nodes and adsorbate--surface edges required to
describe the bonding geometry. The node attributes from the gas-phase
adsorbate and clean surface are then merged into a single feature vector
for each node. This strategy limits the DFT calculations required to obtain node features
to clean surfaces and gas-phase adsorbates. It also avoids the need for
DFT relaxation of every adsorbed structure to build the graph, because
the relevant connectivity is inferred from the matrix describing the chemical bonds.

For TSs, which cannot be represented by a single equilibrium adsorbate graph, the graph was constructed as the union of the initial- and final-state graphs, as also done in a previous work \cite{cheula2025}. The corresponding node features were computed as averages over the initial and final states, yielding a representation that captures both sides of the elementary step without requiring expensive TS optimization methods like nudged elastic band (NEB) \cite{henkelman2000neb,henkelman2000cineb} to define the TS graph. Because this representation uses information from the initial and final states, their formation energies were also included as node features in TS energy prediction, in analogy with BEP relationships.

\begin{figure*}[h]
\centering
\includegraphics[width=0.80\textwidth]{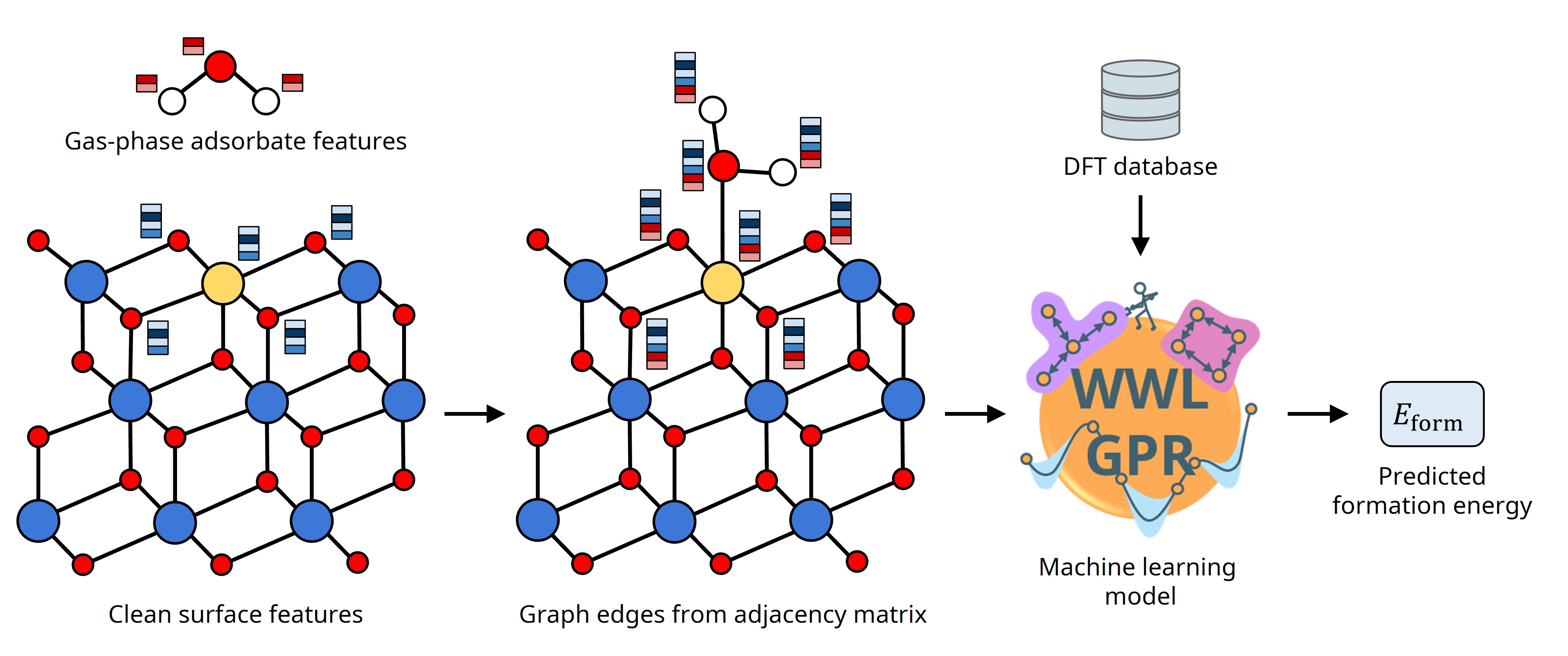}
\caption{
WWL-GPR workflow for predicting formation energies. Gas-phase 
adsorbate features, clean-surface features, and graph edges from an 
adjacency matrix are combined in the WWL-GPR model to predict 
\(E_\mathrm{form}\).
}
\label{fig:workflow}
\end{figure*}

Additional node features employed for both adsorption and TS energy prediction include physical constants~\cite{mendeleev2014}, Bader charges~\cite{henkelman2006bader}, projected density-of-states (pDOS) features, HOMO and LUMO descriptors, and smooth overlap of atomic positions (SOAP) descriptors~\cite{Himanen2020}.
SOAP descriptors provide symmetry-adapted information on the
local atomic environment \cite{bartok2013representing,bartok2017unifies}.
The methodologies used to calculate these features are described in section~S2 of the Supplementary Information.
Importantly, all descriptors used for prediction are derived exclusively 
from clean surfaces and gas-phase adsorbates, rather than 
from the target adsorbed or TS structures, thereby preventing 
information leakage into the regression model.
The WWL kernel defines a similarity measure between attributed
adsorbate--surface graphs by comparing the distributions of node embeddings
obtained after Weisfeiler--Lehman message passing. Gaussian process regression
then uses the resulting graph-kernel matrix to predict formation-energy labels.
\cite{rasmussen2006gpml,shervashidze2011weisfeiler,togninalli2019wasserstein}.
Figure~\ref{fig:workflow} summarizes the complete graph-construction and
prediction workflow.

The feature vectors used in the WWL-GPR model are preprocessed by applying
sample-wise normalization followed by feature-wise min--max scaling to the
interval $[0,1]$. Then, the relative importance of different descriptor classes is
controlled by feature-group weights, which multiply all scaled features within
a given group by the corresponding weight factor.
More details on the features used in the WWL-GPR model are provided in 
Table~S3.
The WWL-GPR hyperparameters are given in Table~S4. The sensitivity of the model predictions to the 
hyperparameters is reported in Figure~S1.

\subsection{Validation procedures}
\label{validation-procedures}

\begin{figure}[h]
\centering
\includegraphics[width=0.40\textwidth]{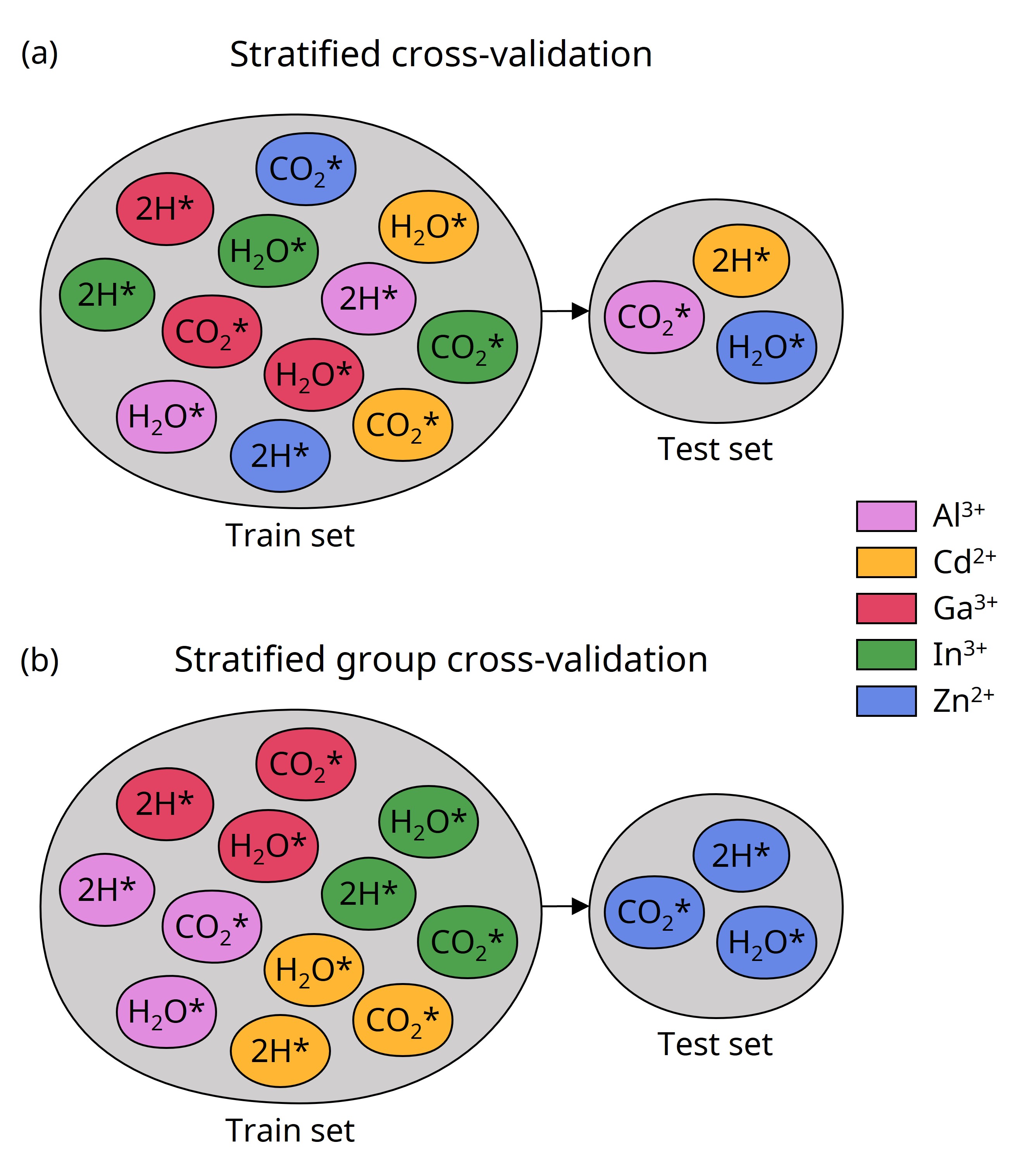}
\caption{
Schematic comparison of (a) stratified cross-validation and (b) stratified 
group cross-validation for doped ZrO$_2$(101) data sets. Colors denote dopants 
Al$^{3+}$, Cd$^{2+}$, Ga$^{3+}$, In$^{3+}$, and Zn$^{2+}$.
}
\label{fig:cv-splits}
\end{figure}

Two validation strategies were used to assess interpolation and dopant
transferability, as described in Figure~\ref{fig:cv-splits}. In stratified
cross-validation, the test set was sampled across both dopant and
adsorbate classes, so that each fold contained chemical environments
represented in the training data. This split evaluates interpolation
within the sampled dopant--adsorbate space. In stratified group
cross-validation, all structures associated with one dopant were held
out from training (as illustrated for Zn$^{2+}$ in Figure~\ref{fig:cv-splits}.b),
except for reference species needed in TRS and WWL-GPR models (see below).
This grouped split provides a stricter transferability test because the model 
must predict energetics for a dopant class that is absent (except for the reference 
species) from the training data.

Prediction performance was quantified using the mean absolute error
(MAE) and root mean square error (RMSE), both reported in eV. The MAE
measures the typical absolute deviation between predicted and reference
energies and is less sensitive to outliers. The RMSE gives greater weight
to large errors and is therefore useful for identifying cases where a few
severe energetic deviations could disproportionately affect the predicted
catalytic activity.

\begin{figure*}[h]
\centering
\includegraphics[width=0.85\textwidth]{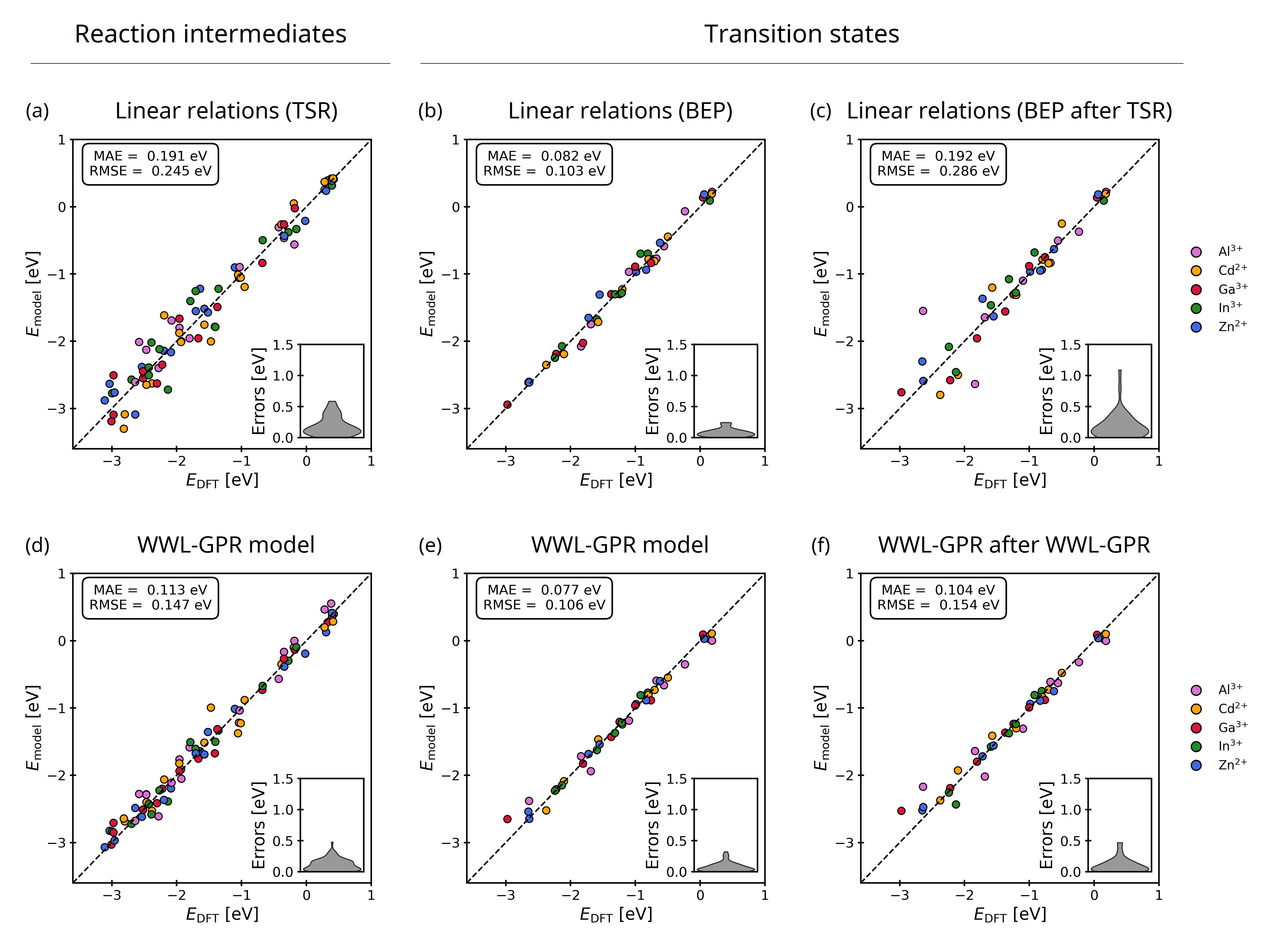}
\caption{
Comparison between DFT and model-predicted formation energies for
reaction intermediates and TSs in stratified cross-validation
(in-domain tasks).
Panels (a,d) show reaction-intermediate energies predicted using TSR and the 
WWL-GPR model, respectively. Panels (b,e) show direct TS predictions
using BEP relations and WWL-GPR, respectively (using DFT-predicted 
intermediate energies). Panels (c,f) show sequential 
TS predictions obtained by combining predicted reaction-intermediate energies with BEP relations or with WWL-GPR, thereby propagating errors in predicted 
reaction-intermediate energies to predicted TS energies.
Inset violin plots show the distributions of absolute prediction errors.
}
\label{fig:stratified-cv}
\end{figure*}

The TSR baselines were constructed by fitting linear regression
models to a common set of reference adsorbate energies, namely 2H*,
OH*+H*, H$_2$O*, and HCOO*+H*. The same reference structures 
were also included in 
the WWL-GPR training data, ensuring that the graph-kernel model and the 
linear baselines were provided with comparable energetic information. 
This is particularly important for dopant-group cross-validation, where 
the shared reference structures provide the model with
information on how simple, key adsorbates bind to the atoms exposed at the doped ZrO$_2$(101) surface. 

\subsection{Activity prediction and rate-control analysis}
\label{activity-prediction-and-rate-control-analysis}

Predicted energetic inputs were propagated through a previously developed microkinetic model 
\cite{cheula2024dopants}
to obtain CH$_3$OH-yield estimates, which were then compared
with yields obtained from DFT-derived energetics. 
The comparison was performed at three reaction conditions: 300 °C and 2 MPa, 340°C 
and 4 MPa, and 380°C and 6 MPa. Across the available dopants and
reaction conditions, the DFT-derived yield spans approximately 0--8\%.
Following our previous work \cite{cheula2024dopants}, the experimental reactor 
of Wang et al. \cite{wang2017znozro2} was modeled as an ideal plug-flow reactor 
(PFR) under purely kinetic conditions, discretized into a series of control 
volumes and integrated by finite differences using the Cantera library \cite{cantera}.
Degree-of-rate-control \cite{campbell2001ratecontrol,stegelmann2009degree}
analysis was used to identify the elementary
steps to which the predicted activity is most sensitive. This analysis
is essential because the usefulness of an energy surrogate is not
determined by its global MAE alone. A model with moderate overall error
can still provide reliable activity predictions if it is accurate for
the kinetically controlling intermediates and TSs. In
contrast, a model with a low global error can give misleading activity
trends if it mispredicts an elementary step with high rate control.

\section{Results and discussion}
\label{results-and-discussion}

\begin{figure*}[h]
\centering
\includegraphics[width=0.85\textwidth]{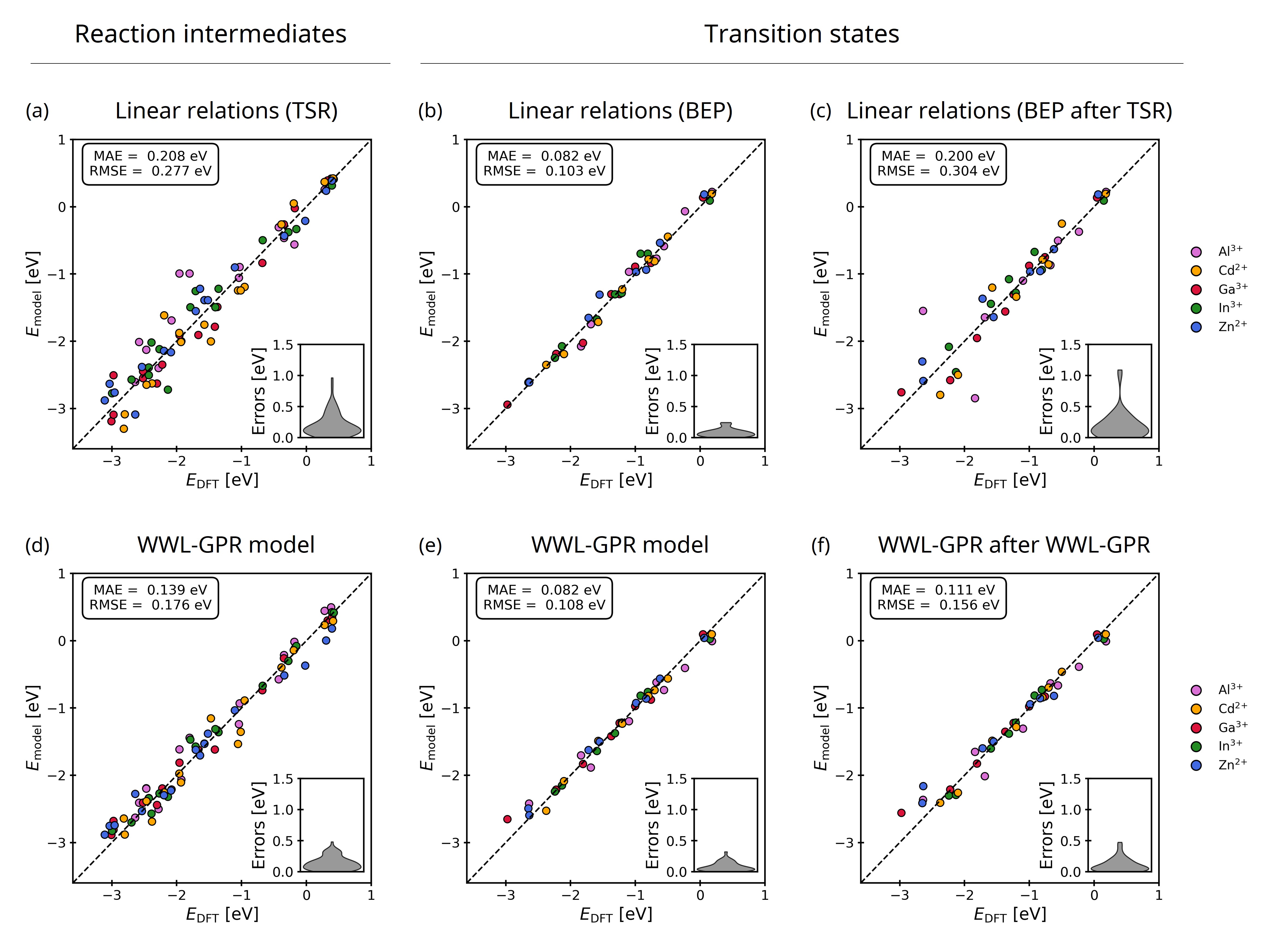}
\caption{
Comparison between DFT and model-predicted formation energies for
reaction intermediates and TSs in stratified dopant-group cross-validation
(out-of-domain tasks).
Panels (a,d) show reaction-intermediate energies predicted using TSR and the 
WWL-GPR model, respectively. Panels (b,e) show direct TS predictions
using BEP relations and WWL-GPR, respectively (using DFT-predicted 
intermediate energies). Panels (c,f) show sequential 
TS predictions obtained by combining intermediate-energy predictions
with BEP relations or with WWL-GPR, thereby propagating errors in predicted 
intermediate energies to predicted TS energies.
Inset violin plots show the distributions of absolute prediction errors.
}
\label{fig:group-cv}
\end{figure*}

\subsection{Stratified cross-validation}
\label{stratified-cross-validation}

Stratified cross-validation evaluates the interpolation performance of
the models when each test fold contains samples from the same dopant and
adsorbate classes represented during training. Under this validation
strategy, TSR provides a baseline description of reaction-intermediate
formation energies, with an MAE of 0.191 eV 
(Figure~\ref{fig:stratified-cv}.a), while
WWL-GPR improves these predictions, reducing the MAE to 0.113 eV 
(Figure~\ref{fig:stratified-cv}.d). Thus, replacing 
the linear TSR model with the
graph-kernel model decreases the errors by approximately 40\%. This
shows that WWL-GPR captures variations in local dopant chemistry and 
adsorbate--surface bonding that are not well-described by a single linear 
scaling relation.

For direct TS prediction, BEP and WWL-GPR give comparable
errors. In this direct task, the initial- and final-state energies used
as descriptors are taken from the DFT reference data. BEP gives an MAE of
0.082 eV (Figure~\ref{fig:stratified-cv}.b), 
whereas WWL-GPR gives an MAE of 0.077 eV
(Figure~\ref{fig:stratified-cv}.e). The small difference between the two
models indicates that WWL-GPR does not provide a clear advantage over BEP
when TS energies are predicted from DFT-quality endpoint
energetics. For the present data set, the TS energies are
therefore already well described by linear energy relationships when
accurate initial- and final-state energies are available.

The benefit of WWL-GPR becomes clearer in the sequential prediction task,
where TS energies are estimated after the initial and final state
energies have also been predicted by the surrogate models. This setting
more closely reflects a screening workflow, in which DFT reference
energies are not available for every reaction intermediate. 
In this case, BEP-after-TSR workflow gives an MAE of 0.200 eV,
(Figure~\ref{fig:stratified-cv}.b), while using WWL-GPR for both adsorbates 
and TS predictions reduces the MAE to 0.104 eV
(Figure~\ref{fig:stratified-cv}.c). 
This corresponds to an error reduction of approximately 50\%, which shows that
WWL-GPR limits the propagation of adsorbate-energy errors into
downstream TS estimates. This is important for catalyst
screening because accumulated energetic errors can alter the predicted
activity trends, even when individual model errors appear moderate.

\subsection{Dopant-group cross-validation}
\label{dopant-group-cross-validation}

The dopant-group cross-validation evaluates transfer to a target dopant that is absent
from the training set, apart from selected reference adsorbate energies
included as described in the Methods section. This validation strategy is
more demanding than stratified cross-validation because the model must
predict adsorption and TS energetics with limited information
on the chemical environment of the target dopant.
For reaction-intermediate energies, TSR provides a baseline description
with an MAE of 0.208 eV
(Figure~\ref{fig:group-cv}.a).
WWL-GPR improves these predictions, reducing the MAE to 0.139 eV
(Figure~\ref{fig:group-cv}.d), which corresponds to an error reduction of
approximately 33\%. The larger WWL-GPR error relative to the
stratified split, where the MAE was 0.113 eV, confirms that dopant
transfer is more challenging than interpolation. Nevertheless, WWL-GPR
retains a clear advantage over TSR, indicating that the attributed graph
representation captures transferable adsorbate--surface and
dopant-dependent information beyond a global linear scaling relation.

For direct TS prediction, BEP and WWL-GPR again give
comparable errors. In this direct task, the initial- and final-state
energies used as descriptors are taken from the DFT reference data. Both BEP and
WWL-GPR give an MAE of 0.082 eV (Figure~\ref{fig:group-cv}.b, and 
Figure~\ref{fig:group-cv}.e). As in the
stratified split, WWL-GPR does not provide a clear advantage over BEP when
TS energies are predicted from DFT-quality endpoint
energetics.
The main improvement of WWL-GPR becomes again clearer in the sequential 
prediction task, where TS energies are estimated after the 
initial- and final-state energies have also been predicted by surrogate models. 
In this case, the BEP-after-TSR workflow gives an MAE of 0.200 eV 
(Figure~\ref{fig:group-cv}.c), while WWL-GPR reduces the MAE to
0.111 eV (Figure~\ref{fig:group-cv}.f). This
corresponds to an error reduction of approximately 45\%, showing that
WWL-GPR limits the propagation of intermediate-energy errors into
downstream TS estimates also in the case of the dopant-group split. 
These results indicate that the primary value of WWL-GPR is not the direct replacement
of BEP when DFT-quality endpoint energies are available, but rather the more accurate
prediction of reaction-intermediate energies and reduced error propagation in
multistep catalyst-screening workflows.

The overall error metrics are summarized in Table~S5 
for stratified cross-validation and in Table~S6 for
dopant-group cross-validation.

For comparison, Kempen et al. \cite{kempen2026foundational} evaluated the zero-shot performance of universal machine-learning potentials (uMLPs) on the same dataset. Their performance varies substantially across models, with some exhibiting errors of several eV. Among the tested uMLPs, eSEN \cite{passaro2023reducing} trained on MPTrj \cite{deng2023chgnet,jain2013materials}, sAlex \cite{schmidt2024improving,barrosoluque2024omat24}, and OMat \cite{barrosoluque2024omat24} shows the best performance on both adsorbates and TSs. For adsorbate formation energies, it achieves an RMSE of 0.180 eV, slightly higher than the RMSE obtained with WWL-GPR. For TS formation energies, it yields an RMSE of 0.134 eV, higher than the RMSEs obtained with direct WWL-GPR prediction but lower than the RMSE of the sequential prediction task.
This comparison should, however, be interpreted with caution, as the uMLPs were evaluated through single-point calculations on DFT-relaxed structures. In catalyst-screening workflows, structures would instead be relaxed using the uMLP itself, introducing additional errors associated with geometry optimization. Kempen et al. \cite{kempen2026foundational} showed, for a different dataset, that this can increase the RMSE by a factor of 1.6. Overall, WWL-GPR achieves an accuracy comparable to the best-performing uMLPs, although the two approaches differ substantially in their training strategy and intended scope.

In the following analyses, we use the dopant-group cross-validation energies, with TS energies obtained from the sequential prediction task, as this best represents the design of new catalysts where all energetics must be predicted by surrogate models.

\subsection{Feature ablation analysis}
\label{features-importance}

\begin{figure}[h]
\centering
\includegraphics[width=0.40\textwidth]{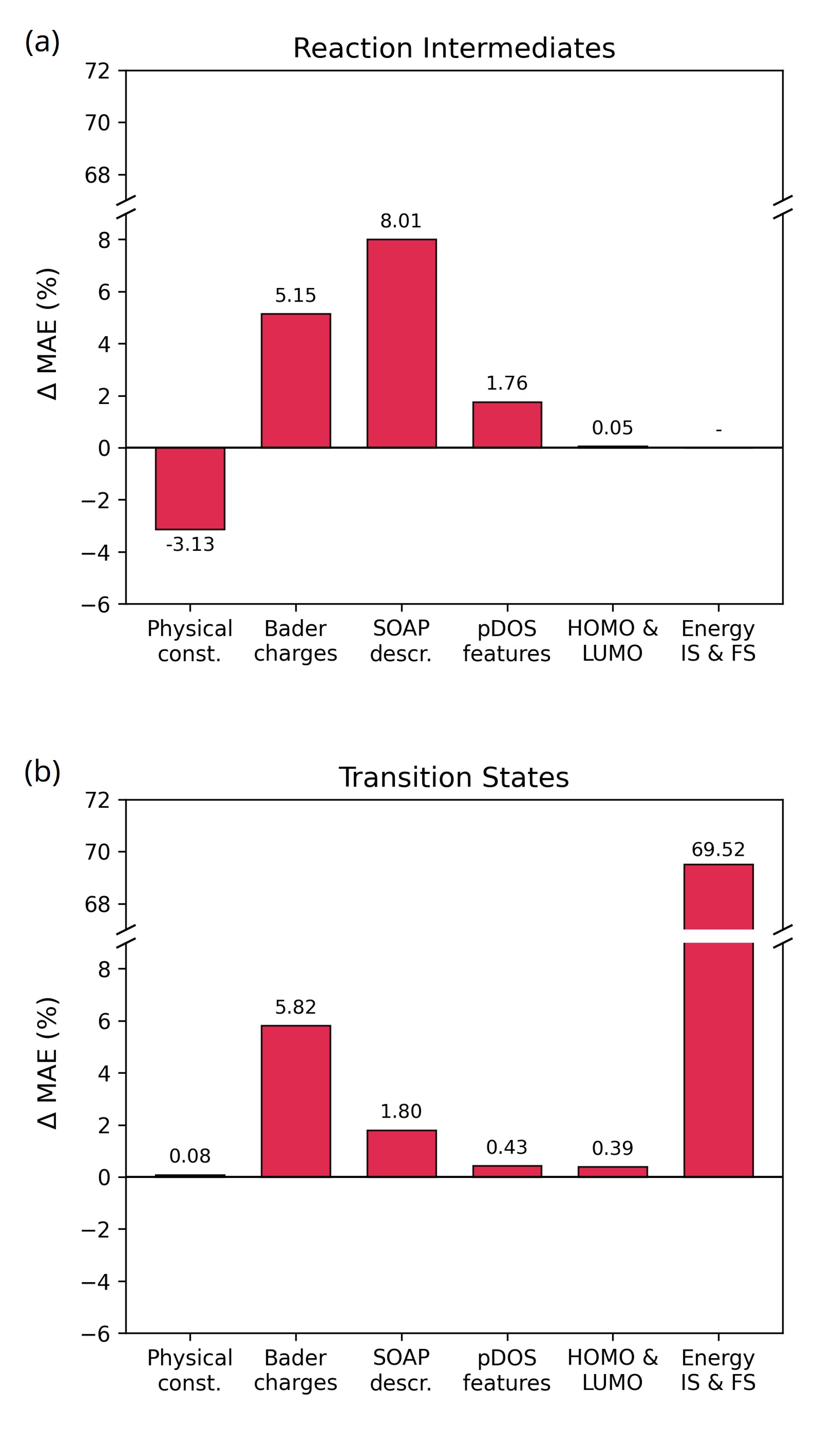}
\caption{
Feature-group ablation analysis for (a) reaction-intermediate and (b)
TS energy predictions. The bars show the percentage change in
MAE obtained when the weight of a selected feature group is reduced to zero
relative to the reference model.
}
\label{fig:feature-ablation}
\end{figure}

To assess the contribution of each feature group in the WWL-GPR model predictions,
we performed an ablation
analysis in which the weight of one feature group at a time was reduced to zero 
while keeping the remaining model settings fixed. The resulting change in MAE,
reported relative to the reference model, is shown in
Figure~\ref{fig:feature-ablation}. Positive values indicate that reducing the
weight of a feature group increases the prediction error, and therefore that
the corresponding descriptors contribute useful information to the model.
Negative values indicate that reducing the weight reduces the error,
suggesting that the feature group is not beneficial for that specific task.

For reaction-intermediate energies, the largest error increase is obtained
when the SOAP descriptors are removed, leading to an MAE increase of
approximately 8\%. Bader charges also contribute substantially with an error
increase of about 5\%, while pDOS features have a smaller but positive effect.
In contrast, HOMO/LUMO descriptors have almost no effect on reaction-intermediate
energy predictions, and the physical-constant features slightly decrease the
MAE when removed. These results indicate that the energy predictions
are mainly controlled by local structural information and charge redistribution,
whereas gas-phase frontier-orbital descriptors play a minor role.

\begin{figure*}[h]
\centering
\includegraphics[width=0.90\textwidth]{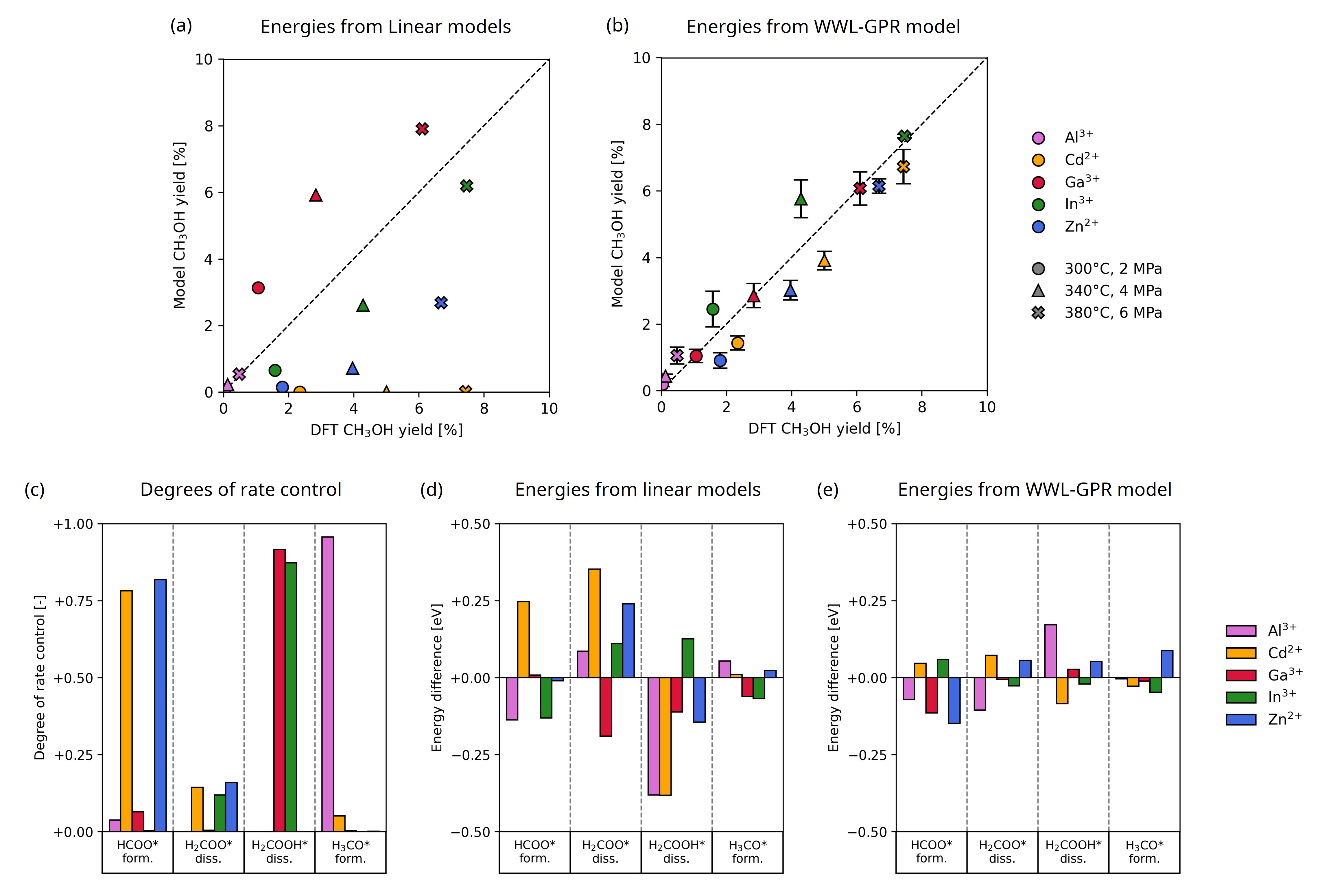}
\caption{
Propagation of surrogate-model energy predictions to catalytic activity.
Panels (a,b) show CH$_3$OH yields obtained from microkinetic simulations using energies
predicted by the linear TSR+BEP workflow (a) and by WWL-GPR (b), compared
with yields obtained from DFT-derived energetics. Colors indicate the dopant
and marker shapes indicate the reaction conditions. The dashed line denotes
parity. Error bars in (b) represent the standard deviations of the predicted yields
obtained from an ensemble of microkinetic models constructed by subsampling 
the WWL-GPR training data.
(c) Degrees of rate control calculated using the DFT-derived energetics at
300~$^\circ$C and 2~MPa for the elementary steps with the largest
rate control. Panels (d,e) show differences between predicted and DFT energies for these
kinetically relevant steps using the linear TSR+BEP workflow (d) and WWL-GPR
(e), respectively.
}
\label{fig:activity}
\end{figure*}

For TS energies, the dominant contribution comes from the
initial- and final-state energies, whose removal increases the MAE by nearly
70\%. This confirms that TS energies are strongly constrained by
the energetics of the connected minima, consistent with the success of
BEP-type relationships for the direct TS prediction task.
Among the remaining descriptors, Bader charges provide the largest contribution,
whereas SOAP, pDOS, HOMO/LUMO, and physical-constant features have smaller
effects. Overall, the ablation analysis shows that different feature groups are
important for intermediates and TSs: local structural descriptors
are most relevant for adsorption energetics, while endpoint energies dominate
TS predictions.

We also performed feature-weight sensitivity tests, in which each feature-group
weight was varied systematically and the resulting MAE was recomputed. These
results are reported in Figure~S2 and 
Figure~S3. The trends are consistent
with the ablation analysis: reaction-intermediate energy predictions are most sensitive
to SOAP descriptors and Bader charges, whereas TS predictions are
dominated by the initial- and final-state energies.

\subsection{Catalytic activity prediction}
\label{catalytic-activity-prediction}

After evaluating the energy-prediction performance of the linear models and
WWL-GPR, we next examine how errors introduced by these surrogate models
propagate to catalytic observables. To this end, the predicted energetics are
used as input to a microkinetic model and the resulting CH$_3$OH yields are
compared with those obtained using DFT-derived energetics.

Energies obtained from the linear TSR+BEP workflow lead to CH$_3$OH-yield
estimates that deviate substantially from the DFT-derived parity line
(Figure~\ref{fig:activity}a). In contrast, yields obtained from WWL-GPR-derived
energies remain closer to the DFT-derived values and better preserve the 
dopant-dependent activity trends (Figure~\ref{fig:activity}b). The uncertainty
in the WWL-GPR-derived yields was estimated using an ensemble of microkinetic
models obtained by subsampling the training data in the cross-validation procedure,
following the approach of Cheula et al.~\cite{cheula2024dopants}. 
The resulting standard deviations of predicted yields are reported as error bars in 
Figure~\ref{fig:activity}b.

This activity-level benchmark is more demanding than an energy-only comparison.
Reaction rates depend exponentially on activation energies, such that moderate
errors in kinetically relevant intermediates or TSs can produce large deviations
in predicted activity. The improved agreement obtained with WWL-GPR therefore
shows that its energetic predictions translate more reliably to the
microkinetic level than those obtained from the linear TSR+BEP workflow.
Nevertheless, the remaining deviations emphasize that even comparatively
accurate energetic surrogates cannot be expected to reproduce DFT-derived
activities exactly.

To identify the elementary steps that most strongly affect the catalytic
activity, we performed a degree-of-rate-control (DRC) analysis,
using the DFT-derived energetics.
Figure~\ref{fig:activity}c reports the DRC values at 300~$^\circ$C and 2~MPa.
The formulas of the mentioned elementary reactions are reported in 
Table~S2.
For Cd$^{2+}$ and Zn$^{2+}$, HCOO* formation exibits the highest DRC, followed
by H$_2$COO* dissociation. H$_2$COOH* dissociation is particularly important 
for Ga$^{3+}$ and In$^{3+}$, while H$_3$CO* formation has the highest DRC
for Al$^{3+}$.
The same qualitative DRC trends are retained at the other reaction conditions
considered (Figure~S4).

The DRC analysis was also repeated using the energetic landscapes predicted
by WWL-GPR and by the linear models (Figure~S5). 
The WWL-GPR energetics better reproduce the DFT-based kinetic picture. For
Cd$^{2+}$ and Zn$^{2+}$, HCOO* formation and H$_2$COO* dissociation remain
the two dominant steps, although their relative importance is reversed, with
H$_2$COO* dissociation showing the larger DRC.
For Ga$^{3+}$ and In$^{3+}$, WWL-GPR correctly retains the importance of 
H$_2$COOH* dissociation, while H$_3$CO* formation remains the dominant step 
for Al$^{3+}$.
In contrast, the linear TSR+BEP workflow deviates more strongly from the
DFT-based DRC analysis, overestimating HCOO* formation for Ga$^{3+}$,
underestimating it for Zn$^{2+}$, and assigning an artificial contribution
to H$_3$CO* formation for Cd$^{2+}$. Thus, large energy errors can alter both
the identity and relative importance of the rate-controlling steps.
Reaction-path analysis further highlights the differences between the models (Figure~S6). While WWL-GPR largely preserves the pathway preferences predicted from DFT, the linear models substantially overestimate the contribution of the H$_2$COOH* dissociation pathway for Cd$^{2+}$ and Zn$^{2+}$ and underestimate the H$_2$COO* dissociation pathway for these dopants.

The comparison of the energy errors for the most kinetically relevant steps
provides a consistent picture (Figure~\ref{fig:activity}d,e). WWL-GPR
generally yields smaller deviations from the DFT energetics than the linear
models for these steps, allowing it to better retain the kinetic structure of
the DFT-based microkinetic model. Even when the relative ordering of the two
dominant DRC contributions is not reproduced exactly, as for Cd$^{2+}$ and
Zn$^{2+}$ (Figure~S5.c), the WWL-GPR model still captures 
the correct subset of kinetically relevant elementary steps and provides a more 
accurate activity prediction.
In general, larger errors in the energetics of kinetically relevant steps lead
to larger deviations in the predicted CH$_3$OH yields.
An exception is observed for Zn$^{2+}$, for which WWL-GPR gives a
larger error for HCOO* formation than the linear TSR+BEP approach, while still 
predicting the CH$_3$OH yield more accurately. 
This can be rationalized by the DRC analysis (Figure~S5.b), 
as the linear workflow significantly underestimates the DRC of HCOO* formation 
while overestimating that of H$_2$COO* dissociation.
This highlights a limitation of DRC analysis, which is a local linear-response
measure around the DFT reference state and therefore cannot fully capture the
combined effect of finite errors in multiple energetic quantities.
When the energetic errors become sufficiently large, they can alter the predicted
DRC values and give incorrect mechanistic understanding and activity trends.
This further emphasizes the importance of the more accurate energetics provided 
by the WWL-GPR workflow.

Taken together, these results show that inaccurate energetic surrogates can affect not only predicted catalytic activities, but also the mechanistic interpretation of the reaction by changing the apparent rate-controlling steps and even the dominant reaction pathway. This has implications beyond catalyst screening, since poor energy models may lead to incorrect conclusions about how the catalytic reaction proceeds. These observations motivate a rate-control-aware active-learning strategy for microkinetic surrogate modeling, in which new DFT calculations are prioritized not only according to predictive uncertainty or their expected contribution to reducing the global MAE, but also according to the kinetic and mechanistic importance of the corresponding elementary steps. Such a criterion could help preserve both activity predictions and mechanistic fidelity, extending active-learning strategies used in catalyst and materials screening \cite{tran2018active,li2021lowcost,tran2021computational,schaaf2023accurate,cheula2026finetuning} to surrogate models coupled directly to microkinetic simulations.

\section{Conclusions}
\label{conclusions}

This work shows that WWL-GPR is a useful surrogate for activity-oriented
screening of doped ZrO$_2$(101) catalysts. Its main strength is
not the replacement of established linear relations in all cases, but the
ability to improve the energetics that most strongly affect downstream
microkinetic predictions. In particular, the graph-kernel representation
provides a more flexible description of dopant-dependent
adsorbate--surface interactions than thermodynamic scaling relations,
while retaining the data efficiency needed for small DFT data sets.

The comparison with BEP relations highlights an important distinction.
When DFT-quality initial- and final-state energies are available, BEP
already provides accurate TS estimates for the present
data set. WWL-GPR becomes more valuable in the sequential screening
setting, where intermediate energies must first be predicted and then
propagated into TS and activity estimates. In this case,
more accurate intermediate energetics reduce error accumulation and lead
to more reliable CH$_3$OH-yield predictions.

The microkinetic analysis further shows that global energy errors provide only an 
incomplete measure of surrogate-model quality. Errors in kinetically important 
intermediates and TSs can have a disproportionate effect on catalytic activity, and 
sufficiently large errors can also alter the predicted rate-controlling steps and dominant
reaction pathways. Poor energetic models can therefore lead not only to inaccurate catalyst
rankings, but also to an incorrect mechanistic interpretation of the catalytic reaction. 
This suggests that future active-learning workflows should
combine model uncertainty with microkinetic sensitivity, so that new DFT
calculations are prioritized according to their expected impact on the
activity prediction rather than on global regression accuracy alone.

\section*{CRediT author contribution statement}
\label{credit-author-contribution-statement}

Karsten Walz Jensen: Conceptualization, Methodology, Investigation,
Writing - original draft.
Mie Andersen: Conceptualization, Supervision, Methodology, Writing -
review and editing. Raffaele Cheula: Conceptualization, Methodology,
Investigation, Supervision, Writing - original draft.

\section*{Declaration of competing interest}
\label{declaration-of-competing-interest}

The authors declare that they have no competing interests.

\section*{Data availability}
\label{data-availability}

The DFT-derived doped-ZrO$_2$(101) data are based on the study
reported by Cheula et al. \cite{cheula2024dopants}. The scripts and data set
to reproduce the results of this work are available on GitHub at
\href{https://github.com/raffaelecheula/doped_ZrO2_modeling}
{\texttt{raffaelecheula/doped\_ZrO2\_modeling}}.

\section*{Acknowledgements}
RC acknowledges support from European Union’s Horizon Europe research and innovation programme under the Marie Skłodowska-Curie grant no.\ 101108769. MA acknowledges funding from Novo Nordisk Fonden (grant no.\ NNF22OC0078939), VILLUM FONDEN (grant no.\ 37381), and the Danish National Research Foundation through the Center of Excellence “InterCat” (grant no.\ DNRF150). Computational support was provided by the Centre for Scientific Computing Aarhus (CSCAA) at Aarhus University.

\bibliographystyle{elsarticle-num}
\bibliography{references}

\end{document}



\onecolumn
\appendix

\renewcommand{\thesection}{S\arabic{section}}
\renewcommand{\thefigure}{S\arabic{figure}}
\renewcommand{\thetable}{S\arabic{table}}
\renewcommand{\theequation}{S\arabic{equation}}

\begin{center}

{\Large \textbf{Supplementary Information}}\\[1.0em]

{\large Improved screening of doped ZrO$_2$ catalysts for CO$_2$ hydrogenation from graph-based machine learning}\\[1.0em]

Karsten Walz Jensen, Mie Andersen$^{*}$, and Raffaele Cheula$^{*}$\\[0.3em]

\textit{Center for Interstellar Catalysis, Department of Physics and Astronomy, Aarhus University, Aarhus C 8000, Denmark}\\[0.3em]

$^{*}$mie@phys.au.dk, $^{*}$raffaele.cheula@phys.au.dk
\end{center}

\section{Reaction mechanism}

\begin{table}[H]
\centering
\caption{Reaction intermediates and corresponding surface configurations.}
\label{tab:intermediates}
\small
\renewcommand{\arraystretch}{1.1}
\begin{tabular}{ccc}
\toprule
& \textbf{Intermediate} & \textbf{Surface configuration} \\
\midrule
1  & 2H*                     & $\mathrm{H^{(M)}{+}H^{(O)}}$ \\
2  & CO$_2$*                 & $\mathrm{CO_2^{(Zr)}}$ \\
3  & HCOO*+H*                & $\mathrm{HCOO^{(Zr,Zr)}{+}H^{(O)}}$ \\
4  & HCOO*+3H*               & $\mathrm{HCOO^{(Zr,Zr)}{+}H^{(M)}{+}H^{(O)}{+}H^{(O)}}$ \\
5  & H$_2$COO*+2H* (on Zr)   & $\mathrm{H_2COO^{(Zr,Zr)}{+}H^{(O)}{+}H^{(O)}}$ \\
6  & H$_2$COO*+2H* (on M)    & $\mathrm{H_2COO^{(M,Zr,Zr)}{+}H^{(O)}{+}H^{(O)}}$ \\
7  & H$_2$COOH*+H*           & $\mathrm{H_2COOH^{(M,Zr,Zr)}+H^{(O)}}$ \\
8  & CH$_2$O*+OH*+H*         & $\mathrm{CH_2O^{(Zr)}{+}OH^{(Zr)}{+}H^{(O)}}$ \\
9  & CH$_2$O*+H$_2$O*        & $\mathrm{CH_2O^{(Zr)}{+}H_2O^{(M)}}$ \\
10 & CH$_2$O*                & $\mathrm{CH_2O^{(Zr)}}$ \\
11 & H$_3$CO+H*              & $\mathrm{H_3CO^{(M)}{+}H^{(O)}}$ \\
12 & H$_3$CO+H* (on Zr)      & $\mathrm{H_3CO^{(Zr)}{+}H^{(O)}}$ \\
13 & CH$_3$OH* (on M)        & $\mathrm{CH_3OH^{(M)}}$ \\
14 & OH*+H* (on Zr)          & $\mathrm{OH^{(Zr)}{+}H^{(O)}}$ \\
15 & OH*+H* (on M)           & $\mathrm{OH^{(M)}{+}H^{(O)}}$ \\
16 & H$_2$O*                 & $\mathrm{H_2O^{(M)}}$ \\
17 & COOH*+H*                & $\mathrm{COOH^{(M,Zr)}{+}H^{(O)}}$ \\
18 & CO*                     & $\mathrm{CO^{(Zr)}}$ \\
19 & 2H* (on Zr)             & $\mathrm{H^{(Zr)}{+}H^{(O)}}$ \\
20 & CO$_3$* (on M)          & $\mathrm{CO_2^{(M,Zr,Zr,O)}}$ \\
21 & CO$_3$* (on Zr)         & $\mathrm{CO_2^{(Zr,Zr,Zr,O)}}$ \\
\bottomrule
\end{tabular}
\end{table}

\begin{table}[H]
\centering
\caption{Elementary reactions included in the microkinetic model.}
\label{tab:reactions}
\small
\renewcommand{\arraystretch}{1.1}
\begin{tabular}{cccc}
\toprule
& \textbf{Reaction} & \textbf{Reaction equation} & \textbf{Activated} \\
\midrule
1 & H$_2$ adsorption
& $\mathrm{H_2 + {}^{(M)} + {}^{(O)} \rightleftharpoons H^{(M)}{+}H^{(O)}}$ & yes \\
2 & CO$_2$ adsorption
& $\mathrm{CO_2 + {}^{(Zr)} \rightleftharpoons CO_2^{(Zr)}}$ & no \\
3 & HCOO* formation
& $\mathrm{CO_2^{(Zr)} + H^{(M)}{+}H^{(O)} + {}^{(Zr)} \rightleftharpoons HCOO^{(Zr,Zr)}{+}H^{(O)} + {}^{(M)}}$ & yes \\
4 & H$_2$COO* formation
& $\mathrm{HCOO^{(Zr,Zr)}{+}H^{(O)} + H^{(M)}{+}H^{(O)} \rightleftharpoons H_2COO^{(M,Zr,Zr)}{+}H^{(O)}{+}H^{(O)}}$ & yes \\
5 & H$_2$COO* dissociation
& $\mathrm{H_2COO^{(M,Zr,Zr)}{+}H^{(O)}{+}H^{(O)} \rightleftharpoons CH_2O + OH^{(Zr)}{+}H^{(O)} + {}^{(M)} + {}^{(Zr)} + {}^{(O)}}$ & yes \\
6 & H$_2$COOH* formation
& $\mathrm{H_2COO^{(M,Zr,Zr)}{+}H^{(O)}{+}H^{(O)} \rightleftharpoons H_2COOH^{(M,Zr,Zr)}{+}H^{(O)} + {}^{(O)}}$ & yes \\
7 & H$_2$COOH* dissociation
& $\mathrm{H_2COOH^{(M,Zr,Zr)}{+}H^{(O)} \rightleftharpoons CH_2O + H_2O^{(M)} + 2\,{}^{(Zr)} + {}^{(O)}}$ & yes \\
8 & H$_3$CO* formation
& $\mathrm{CH_2O + H^{(M)}{+}H^{(O)} \rightleftharpoons H_3CO^{(M)}{+}H^{(O)}}$ & yes \\
9 & CH$_3$OH desorption
& $\mathrm{H_3CO^{(M)}{+}H^{(O)} \rightleftharpoons CH_3OH + {}^{(M)} + {}^{(O)}}$ & yes \\
10 & H$_2$O* formation
& $\mathrm{OH^{(Zr)}{+}H^{(O)} + {}^{(M)} \rightleftharpoons H_2O^{(M)} + {}^{(Zr)} + {}^{(O)}}$ & yes \\
11 & H$_2$O desorption
& $\mathrm{H_2O^{(M)} \rightleftharpoons H_2O + {}^{(M)}}$ & no \\
12 & COOH* dissociation
& $\mathrm{COOH^{(M,Zr)}{+}H^{(O)} \rightleftharpoons CO^{(Zr)} + OH^{(M)} + H^{(O)}}$ & yes \\
13 & CO desorption
& $\mathrm{CO^{(Zr)} \rightleftharpoons CO + {}^{(Zr)}}$ & no \\
14 & H$_2$ adsorption (on Zr)
& $\mathrm{H_2 + {}^{(Zr)} + {}^{(O)} \rightleftharpoons H^{(Zr)}{+}H^{(O)}}$ & yes \\
15 & CO$_3$* formation (on M)
& $\mathrm{CO_2 + {}^{(M)} + 2\,{}^{(Zr)} + {}^{(O)} \rightleftharpoons CO_2^{(M,Zr,Zr,O)}}$ & no \\
16 & CO$_3$* formation (on Zr)
& $\mathrm{CO_2 + 3\,{}^{(Zr)} + {}^{(O)} \rightleftharpoons CO_2^{(Zr,Zr,Zr,O)}}$ & no \\
\bottomrule
\end{tabular}
\end{table}

Tables ~\ref{tab:intermediates} and ~\ref{tab:reactions} summarize the reaction intermediates and the elementary reactions included in the microkinetic model, respectively. The symbols in the superscripts indicate the atoms of the active sites to which the adsorbed species are bonded, with "M" defining the surface dopant atom. Activated reactions are treated using TS theory, whereas non-activated reactions are described using the kinetic theory of gases, as described in Cheula et al. \cite{cheula2024dopants}.

\section{Model features}

Table~\ref{tab:features} summarizes the node features used in the graph representation, including their definitions, units, treatment of missing values (described in more detail below), and feature classes. The feature classes describe how the corresponding quantities are obtained: "atomic" features are intrinsic properties of the atomic species; "site" features are calculated from the DFT structures representing the relaxed clean catalyst surface and unrelaxed corresponding species in the gas phase; "molecule" features are calculated only from the unrelaxed species in the gas phase, with geometries obtained with the CatKit library \cite{Boes2019CatKit}; and "reaction" features are derived from the energies of the initial state (IS) and final state (FS) of the corresponding reaction and are included only for TS species. 
While "atomic" and "site" features are assigned different values for each atom in the graph, "molecule" features are assigned a value for the atoms of the adsorbed species and zero for the atoms of the surface, and "reaction" features are assigned the same value to all atoms in the graph.

\begin{table}[H]
\centering
\caption{Node features used in the graph representation.}
\label{tab:features}
\small
\renewcommand{\arraystretch}{1.15}
\begin{tabular}{cccccc}
\toprule
\textbf{Group} & \textbf{Feature} & \textbf{Description} & \textbf{Unit} & \textbf{Value if missing} & \textbf{Class} \\
\midrule
\multirow{3}{*}{Physical constants}
& $\mathrm{EA}$ & Electron affinity & eV & -- & Atomic \\
& $\chi$ & Electronegativity & eV & -- & Atomic \\
& $\mathrm{IP}$ & Ionization potential & eV & -- & Atomic \\
\midrule
{Bader charges}
& $B_C$ & Bader charge & e & $+0.0$ & Site \\
\midrule
\multirow{9}{*}{pDOS (d band)}
& $f_d$ & d-band filling & e & $-1.3$ & Site \\
& $\epsilon^{\mathrm{unocc}}_d$ & Unoccupied d-band center & eV & $+2.6$ & Site \\
& $W^{\mathrm{unocc}}_d$ & Unoccupied d-band width & eV & $+0.9$ & Site \\
& $\mathrm{Skew}^{\mathrm{unocc}}_d$ & Unoccupied d-band skewness & -- & $+0.9$ & Site \\
& $\mathrm{Kurt}^{\mathrm{unocc}}_d$ & Unoccupied d-band kurtosis & -- & $-1.0$ & Site \\
& $\epsilon^{\mathrm{occ}}_d$ & Occupied d-band center & eV & $-6.3$ & Site \\
& $W^{\mathrm{occ}}_d$ & Occupied d-band width & eV & $+1.0$ & Site \\
& $\mathrm{Skew}^{\mathrm{occ}}_d$ & Occupied d-band skewness & -- & $-0.7$ & Site \\
& $\mathrm{Kurt}^{\mathrm{occ}}_d$ & Occupied d-band kurtosis & -- & $+0.9$ & Site \\
\midrule
\multirow{5}{*}{pDOS (s and p bands)}
& $f_{sp}$ & sp-band filling & e & $-3.5$ & Site \\
& $\epsilon^{\mathrm{occ}}_p$ & Occupied p-band center & eV & $-6.6$ & Site \\
& $W^{\mathrm{occ}}_p$ & Occupied p-band width & eV & $+0.7$ & Site \\
& $\mathrm{Skew}^{\mathrm{occ}}_p$ & Occupied p-band skewness & -- & $-1.9$ & Site \\
& $\mathrm{Kurt}^{\mathrm{occ}}_p$ & Occupied p-band kurtosis & -- & $+0.5$ & Site \\
\midrule
{SOAP descriptor}
& $\mathrm{SOAP}$ & Smooth overlap of atomic positions & -- & -- & Site \\
\midrule
\multirow{2}{*}{HOMO and LUMO}
& $\mathrm{HOMO}$ & Highest occupied molecular orbital & eV & $+0.0$ & Molecule \\
& $\mathrm{LUMO}$ & Lowest unoccupied molecular orbital & eV & $+0.0$ & Molecule \\
\midrule
\multirow{2}{*}{IS and FS energies}
& $E_{\mathrm{IS}}$ & Initial-state energy & eV & -- & Reaction \\
& $E_{\mathrm{FS}}$ & Final-state energy & eV & -- & Reaction \\
\bottomrule
\end{tabular}
\end{table}

Physical constants of the atomic species were taken from the tabulated data available in the Mendeleev Python library~\cite{mendeleev2014}.
Bader charge analysis was performed using the grid-based algorithm developed by Henkelman and co-workers~\cite{henkelman2006bader}. 
SOAP descriptors for the clean surfaces and unrelaxed gas-phase molecules were calculated using DScribe~\cite{Himanen2020}, with a cutoff radius of 5.0~\AA, a Gaussian smearing width of 0.5~\AA, and radial and angular basis truncations of $n_{\max}=3$ and $l_{\max}=2$, respectively. No distinction between chemical species was made in the SOAP calculation, resulting in a feature vector comprising 18 SOAP components.

All pDOS calculations were performed using the settings outlined in Cheula et al.~\cite{cheula2024dopants}, but with a tighter $k$-point grid of $14 \times 14 \times 1$, together with the smearing-free tetrahedron method and an energy spacing of 0.01~eV. For the occupied-state pDOS features, the energy ranges used in the formulas below were restricted to $-10.0$ to $0.0$~eV relative to the Fermi level. This restriction was introduced because sharp states were observed at lower energies and were assumed to contribute only marginally to the bonding of the different intermediates. All calculated spectra showed characteristics similar to those reported for pristine ZrO$_2$ by Zhang et al.~\cite{ref_pdos_ZHANG}.

All features are calculated from the clean surfaces and gas-phase of the adsorbates. Hence, we have cases where features cannot be meaningfully defined. For adsorbates, we cannot define any pDOS-related features or Bader charges, as they are calculated only for the clean surfaces. Furthermore, some atoms in the surfaces do not have $d$ states, meaning we cannot assign any $d$-band moments. Finally, the HOMO/LUMO features were obtained from the electronic structure of the corresponding unrelaxed gas-phase species. Here we also have undefined values, as they are derived from the molecular phase of the adsorbate, and hence not defined for any atoms within the surface. In all of these cases, we assign a "missing" value (see Table~\ref{tab:features}). For HOMO/LUMO and Bader features, we use a value of zero to distinguish this case from the positive or negative values in the data set. For the other features, the value was defined from its distribution in the data set as the minimum observed value minus two standard deviations, ensuring that it lies well outside the range of physically meaningful values. The resulting value was then assigned to all nodes corresponding to atoms whenever a feature cannot be defined.

The formulas used for calculating the different band moments are listed below:
\begin{align*}
    \epsilon_{p,d} &= \frac{\int\text{pDOS}(E)\cdot E dE}{\int\text{pDOS}(E)dE} \\
    W_{p,d} &= \sqrt{\frac{\int\text{pDOS}(E)(E-\epsilon_{p,d})^2 dE}{\int\text{pDOS}(E)dE} }\\
    \mathrm{Skew}_{p,d} &= \frac{1}{(W_{p,d})^3}\frac{\int\text{pDOS}(E)(E-\epsilon_{p,d})^3 dE}{\int\text{pDOS}(E)dE}\\
    \mathrm{Kurt}_{p,d} &= \frac{1}{(W_{p,d})^4}\frac{\int\text{pDOS}(E)(E-\epsilon_{p,d})^4 dE}{\int\text{pDOS}(E)dE}\\
\end{align*}

\section{Model parameters}

The WWL-GPR model parameters used throughout this work are summarized in
Table~\ref{tab:model-parameters}. The GPR length scale was optimized within
the interval $[1,1000]$, while a regularization term of $10^{-6}$ was used
to improve numerical stability. Different node weights were assigned to
adsorbate atoms and to atoms in the first and second surface coordination
shells, with progressively lower weights for atoms farther from the
adsorption site. All edge types were assigned equal weight. Among the node
feature groups, physical constants, Bader charges, SOAP descriptors, pDOS
features, and HOMO/LUMO descriptors were weighted equally, whereas the
initial-state and final-state energy descriptors were assigned a larger weight
of 10 to emphasize their known importance for TS energies (i.e., the existence of BEP relations).

\begin{table}[H]
\centering
\caption{Hyperparameters and weights used in the WWL-GPR model.}
\label{tab:model-parameters}
\small
\renewcommand{\arraystretch}{1.15}
\begin{tabular}{ccc}
\toprule
\textbf{Category} & \textbf{Parameter} & \textbf{Value} \\
\midrule
\multirow{2}{*}{GPR parameters} & Length-scale bounds & $[1, 1000]$ \\
& Regularization & $10^{-6}$ \\
\midrule
\multirow{3}{*}{Node weights} & Adsorbates & $1.0$ \\
& First surface shell & $0.9$ \\
& Second surface shell & $0.1$ \\
\midrule
\multirow{3}{*}{Edge weights} & Adsorbate--adsorbate & $1.0$ \\
& Adsorbate--surface & $1.0$ \\
& Surface--surface & $1.0$ \\
\midrule
\multirow{6}{*}{Feature group weights} & Physical constants & $1.0$ \\
& Bader charges & $1.0$ \\
& SOAP descriptor & $1.0$ \\
& pDOS features & $1.0$ \\
& HOMO and LUMO & $1.0$ \\
& IS and FS energies & $10.0$ \\
\bottomrule
\end{tabular}
\end{table}

\section{Model performances}

Tables~\ref{tab:stratified-cv-errors} and  ~\ref{tab:group-cv-errors} summarize the MAE and RMSE obtained for the different models and prediction tasks under stratified and dopant-group cross-validation, respectively.

\begin{table}[H]
\centering
\caption{Stratified cross-validation error summary. Errors are reported in eV.}
\label{tab:stratified-cv-errors}
\small
\begin{tabular}{cccc}
\toprule
\textbf{Task} & \textbf{Model} & \textbf{MAE} & \textbf{RMSE} \\
\midrule
Intermediate energy & TSR & 0.191 & 0.245 \\
Intermediate energy & WWL-GPR & 0.113 & 0.147 \\
Direct transition state & BEP & 0.082 & 0.103 \\
Direct transition state & WWL-GPR & 0.077 & 0.106 \\
Sequential transition state & TSR+BEP & 0.200 & 0.304 \\
Sequential transition state & WWL-GPR & 0.104 & 0.154 \\
\bottomrule
\end{tabular}
\end{table}

\begin{table}[H]
\centering
\caption{Dopant-group cross-validation error summary. Errors are reported in eV.}
\label{tab:group-cv-errors}
\small
\begin{tabular}{cccc}
\toprule
\textbf{Task} & \textbf{Model} & \textbf{MAE} & \textbf{RMSE} \\
\midrule
Intermediate energy & TSR & 0.208 & 0.277 \\
Intermediate energy & WWL-GPR & 0.139 & 0.176 \\
Direct transition state & BEP & 0.082 & 0.103 \\
Direct transition state & WWL-GPR & 0.082 & 0.108 \\
Sequential transition state & TSR+BEP & 0.200 & 0.304 \\
Sequential transition state & WWL-GPR & 0.111 & 0.156 \\
\bottomrule
\end{tabular}
\end{table}

\section{Model parameters and features sensitivity analysis}

\begin{figure}[H]
\centering
\includegraphics[width=\textwidth]{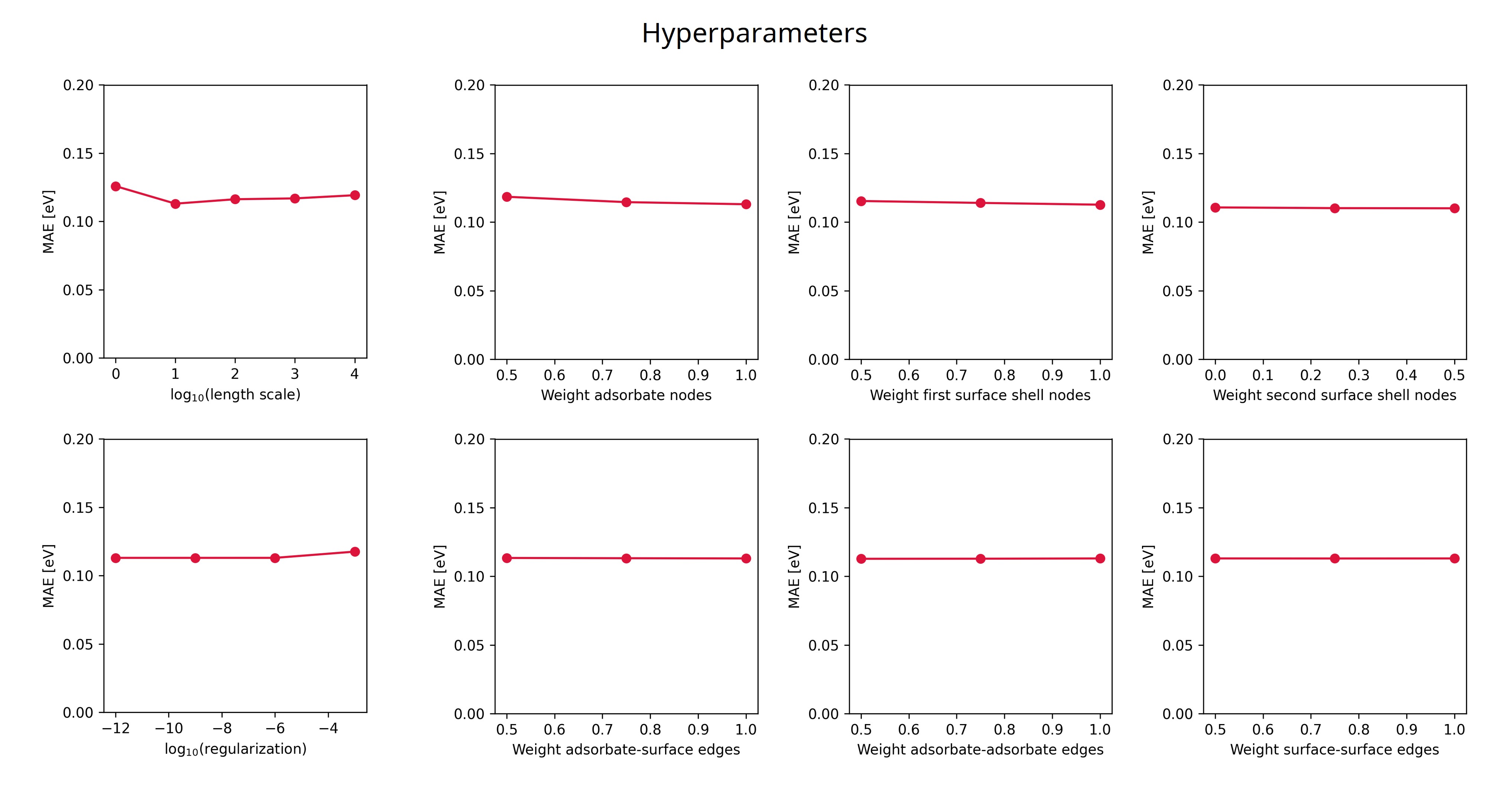}
\caption{
Sensitivity of the WWL-GPR prediction accuracy to the main model
hyperparameters. The MAE is reported as a function of the kernel length scale,
regularization parameter, node weights, and edge weights, with all other model
parameters kept fixed.
}
\label{fig:sensitivity-parameters}
\end{figure}

\begin{figure}[H]
\centering
\includegraphics[width=0.75\textwidth]{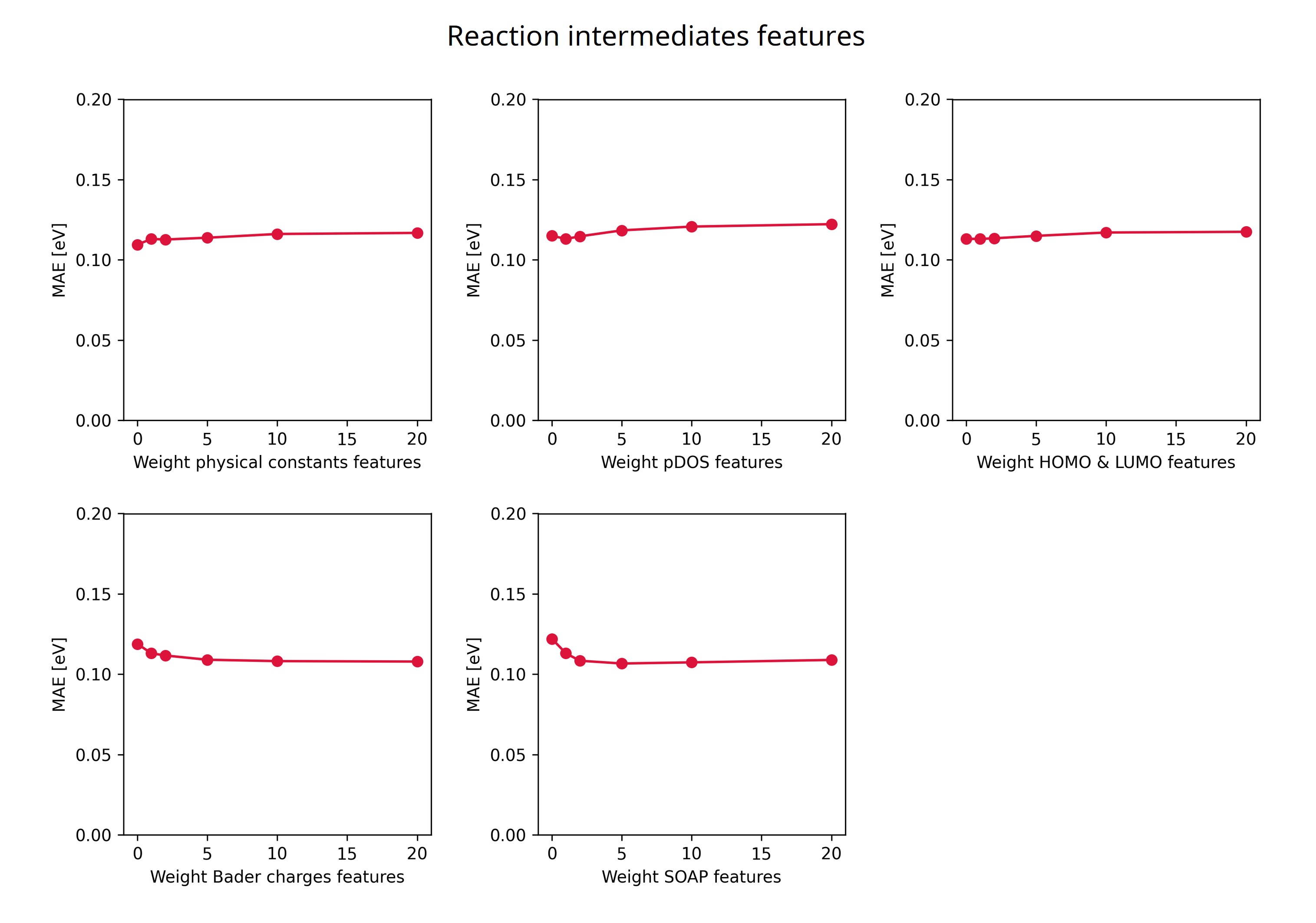}
\caption{
Sensitivity of reaction-intermediate predictions to the weights assigned
to the different feature groups. The MAE is reported as a function of the weight
of physical-constant, pDOS, HOMO/LUMO, Bader-charge, and SOAP features, while
keeping the remaining model parameters fixed.
}
\label{fig:sensitivity-features-intermediates}
\end{figure}

\begin{figure}[H]
\centering
\includegraphics[width=0.75\textwidth]{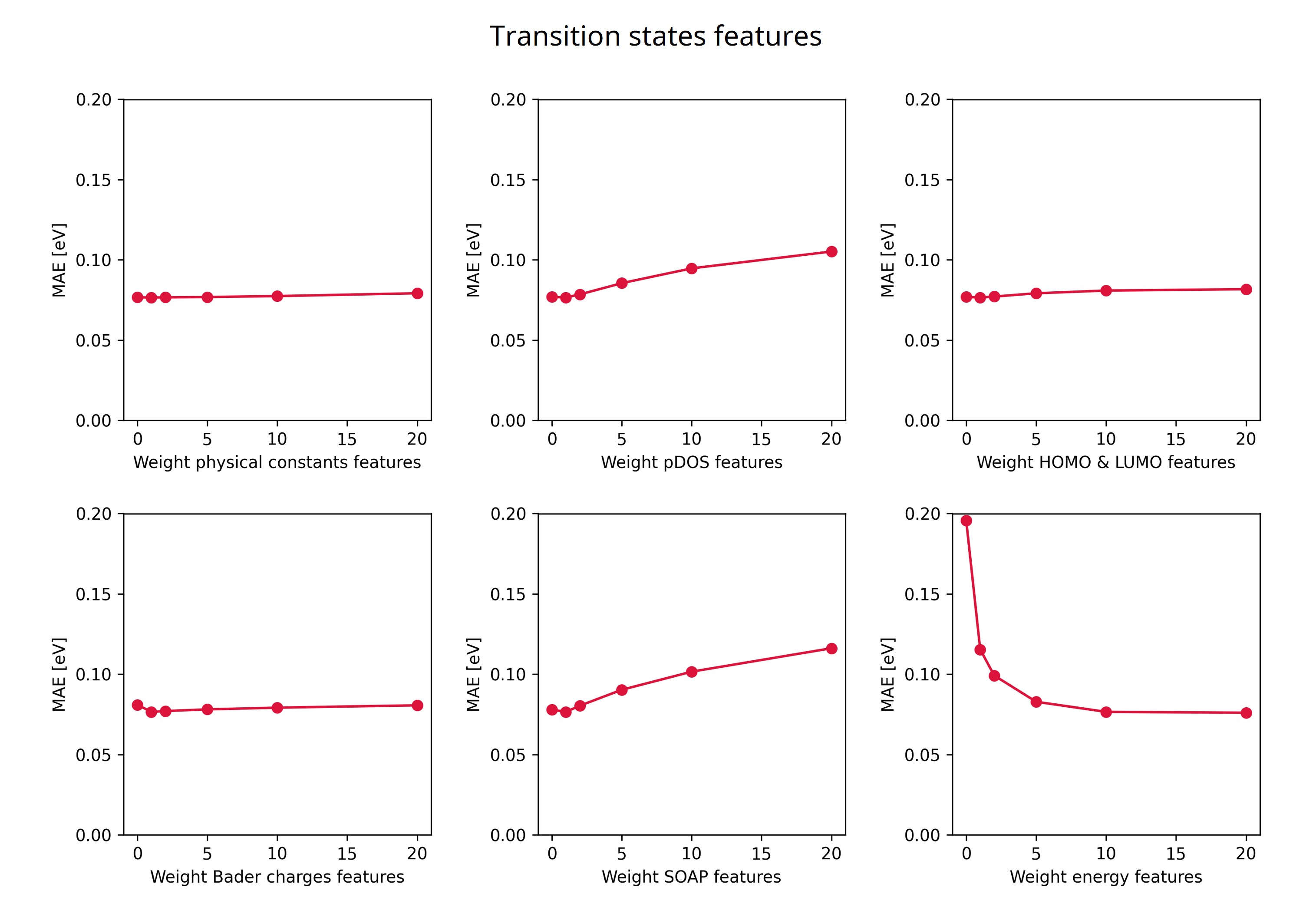}
\caption{
Sensitivity of TS energy predictions to the weights assigned to
the different feature groups. The MAE is reported as a function of the weight
of physical-constant, pDOS, HOMO/LUMO, Bader-charge, SOAP, and initial- and
final-state energy features, while keeping the remaining model parameters fixed.
}
\label{fig:sensitivity-features-reactions}
\end{figure}

The sensitivity of the WWL-GPR model to the main hyperparameters and feature-group
weights is shown in Figs.~\ref{fig:sensitivity-parameters}--\ref{fig:sensitivity-features-reactions}.
The prediction error is relatively insensitive to most node and edge weights and
to the regularization parameter, whereas a mild dependence is observed for the
kernel length scale. Similarly, the weights assigned to most feature groups have
only a moderate effect on the prediction accuracy. The strongest dependence is
observed for TS predictions when varying the weight of the initial- and final-state
energy descriptors, whose inclusion substantially reduces the MAE. These results
indicate that the model performance is generally robust with respect to the chosen
hyperparameters, while reaction-energy information plays a particularly important
role in describing TS energetics.

\section{Microkinetic analysis}

\begin{figure}[H]
\centering
\includegraphics[width=0.95\textwidth]{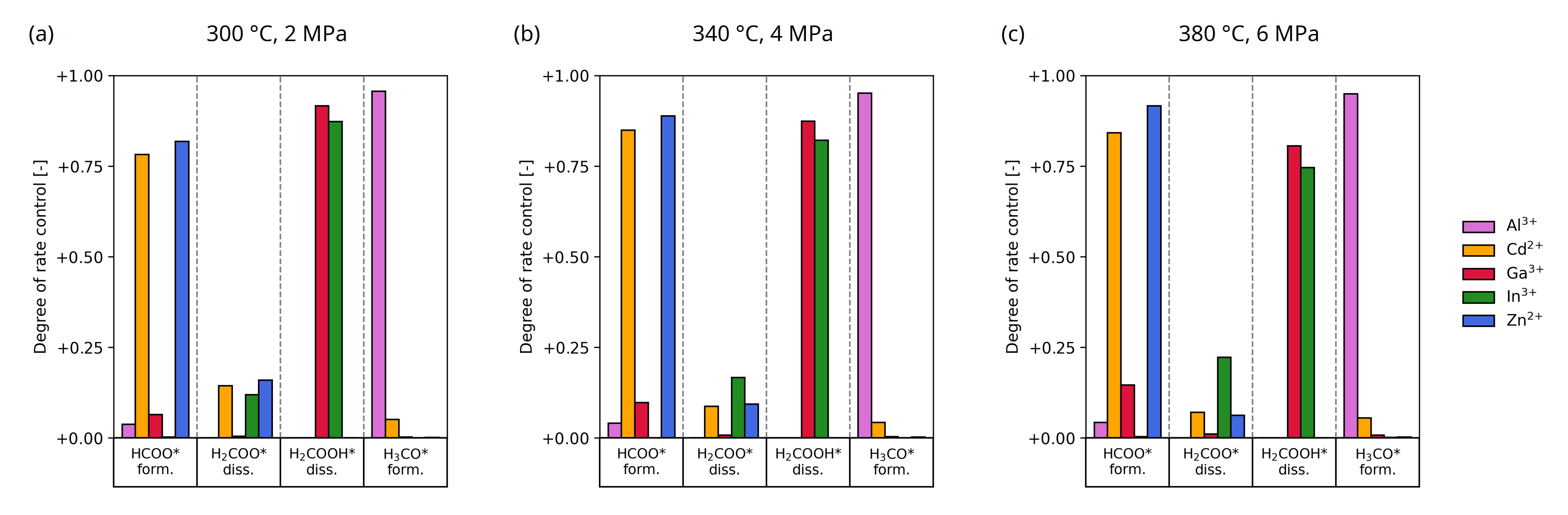}
\caption{
Dependence of the degree of rate control on reaction conditions.
DRC values calculated from the DFT-derived energetics are shown for
(a) 300~$^\circ$C and 2~MPa, (b) 340~$^\circ$C and 4~MPa, and
(c) 380~$^\circ$C and 6~MPa. Colors identify the dopant. Although the
magnitudes vary with reaction conditions, the same elementary steps remain
among the dominant contributors to the activity sensitivity.
}
\label{fig:drc-conditions}
\end{figure}

\begin{figure}[H]
\centering
\includegraphics[width=0.95\textwidth]{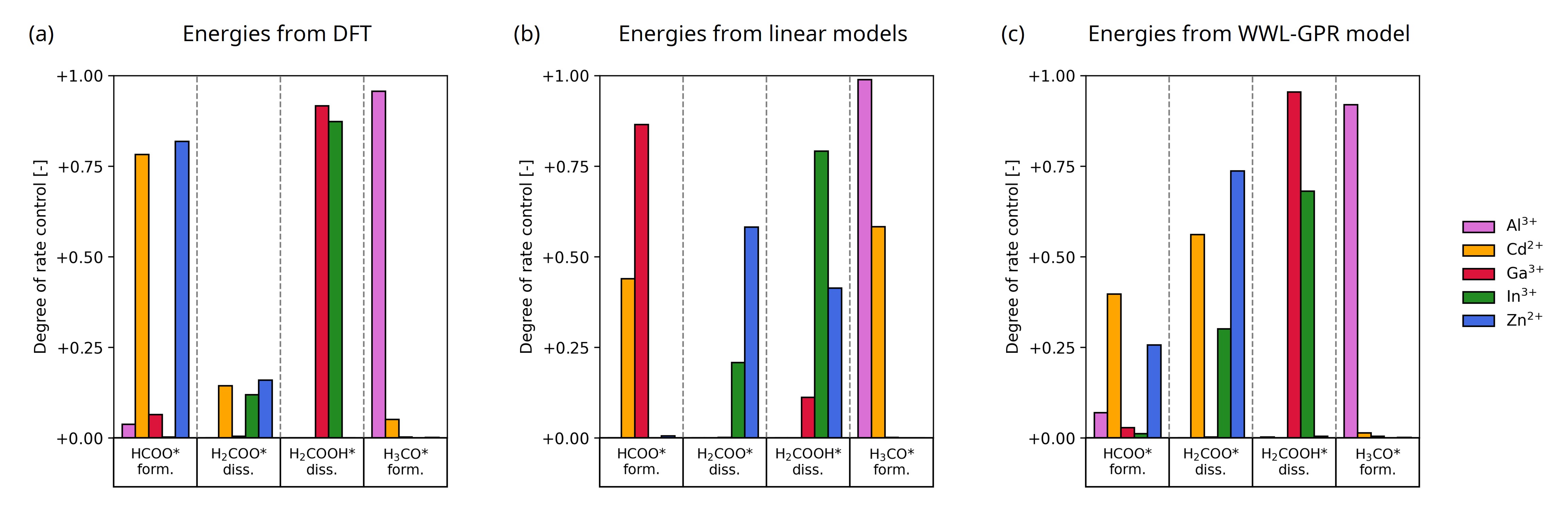}
\caption{
Comparison of degrees of rate control obtained using different energetic
descriptions at 300~$^\circ$C and 2~MPa.
DRC values are calculated using (a) DFT-derived energies,
(b) energies predicted by the linear TSR+BEP workflow, and
(c) energies predicted by WWL-GPR. Colors identify the dopant.
The comparison illustrates how errors in the energetic surrogate can alter
the predicted kinetic sensitivities of the microkinetic model.
}
\label{fig:drc-models}
\end{figure}

\begin{figure}[H]
\centering
\includegraphics[width=0.95\textwidth]{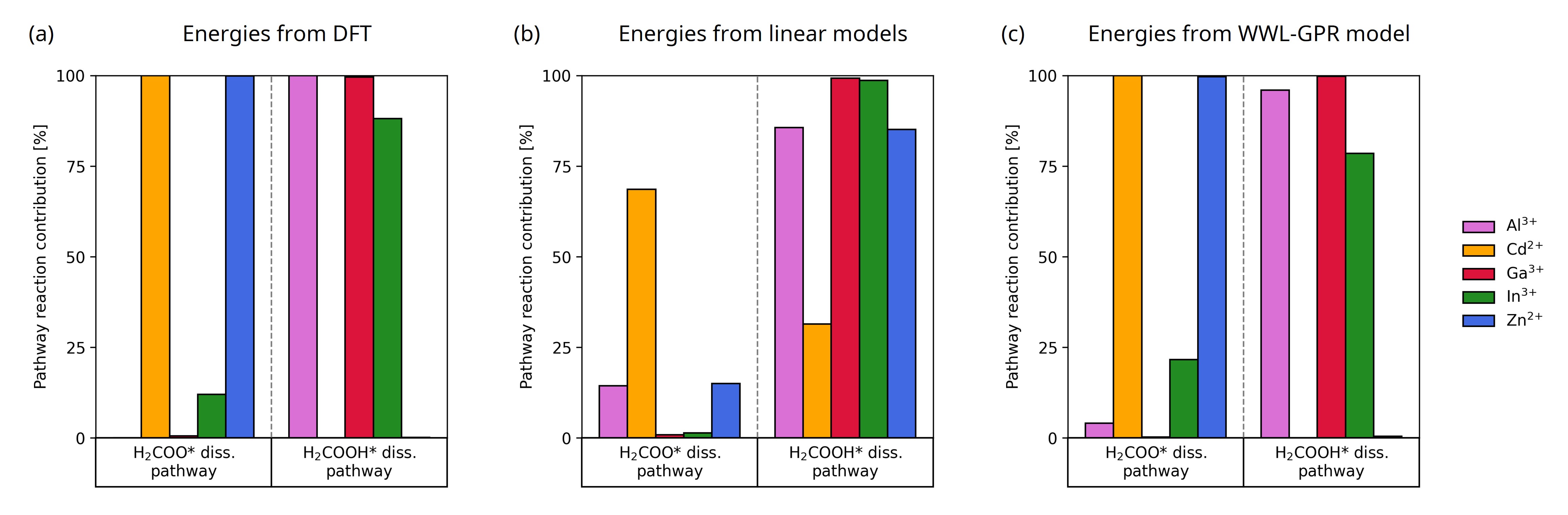}
\caption{
Relative contributions of the H$_2$COO* and H$_2$COOH* dissociation pathways for the different dopants, calculated using (a) DFT energetics, (b) linear-model energetics, and (c) WWL-GPR energetics.
}
\label{fig:pathways}
\end{figure}



\bibliographystyle{elsarticle-num}
\bibliography{references}